\documentstyle[aps,epsfig,prd,floats]{revtex}

\newcommand{\mean}[1]{\langle #1 \rangle}

\newcommand{\eq}[1]{\begin{equation} #1 \end{equation}} 

\newcommand{\non}{$N_{\sl on}$ {}}
\newcommand{\noff}{$N_{\sl off}$ {}}
\newcommand{\lon}{{\sl on}}
\newcommand{\loff}{{\sl off}}

\begin{document}

\draft 

\wideabs{
\title{Correlation between Gamma-Ray bursts  and Gravitational Waves}
\author{P. Tricarico\footnotemark[1], A. Ortolan, A. Solaroli and G. Vedovato}
\address{I.N.F.N., Laboratori Nazionali di Legnaro, via Romea, 4, I-35020 Legnaro, Padova, Italy}
\author{L. Baggio, M. Cerdonio, L. Taffarello and J. Zendri}
\address{Phys. Dept., Univ. of Padova and I.N.F.N. Sez. Padova, Via Marzolo, 8, I-35100 Padova, Italy}
\author{R. Mezzena, G.A. Prodi and S. Vitale}
\address{Phys. Dept., Univ. Trento and I.N.F.N. Gruppo Coll. Trento, Sez. Padova, I-38050 Povo, Italy.}
\author{P. Fortini}
\address{Phys. Dept., University of Ferrara, and INFN Sez. Ferrara, via Paradiso, 12, I-44100 Ferrara, Italy}
\author{M. Bonaldi and P. Falferi}
\address{Centro di Fisica degli Stati Aggregati, I.T.C.-C.N.R., Trento, Italy and \\ I.N.F.N. Gruppo Coll. Trento, Sez. Padova, I-38050 Povo, Italy.}
\date{\today}
\maketitle 
\begin{abstract}%
The cosmological origin of $\gamma$-ray bursts (GRBs) is now commonly accepted and,
according to several models for the central engine, GRB sources should also emit at the same time
gravitational waves bursts (GWBs). We have performed   
two correlation searches between the data of
the resonant gravitational wave detector AURIGA and GRB arrival times collected in the
BATSE 4B catalog.  No correlation was found and  
an upper limit \bbox{$h_{\text{RMS}} \leq 1.5 \times 10^{-18}$} on the averaged amplitude of
gravitational waves associated with $\gamma$-ray bursts has been set for the first time.
\end{abstract}
}

\narrowtext


\footnotetext[1]{Corresponding author, \\ e-mail: {\tt Pasquale.Tricarico@lnl.infn.it}}

\section{Introduction}

Thirty years after their discovery, Gamma Ray Bursts (GRBs) are still
the most mysterious objects in the Universe,
as their properties have not yet been associated to
any well-known object, while for instance pulsars have been almost immediately 
identified as final objects of the evolution of stars and quasars
have been well framed among galaxy nuclei in particular in the Seyfert
class. Unfortunately ``standard'' astrophysical objects do not exist for GRBs.
The ``non standard'' compact astrophysical objects which might reproduce the
experimental characteristic of GRBs involve black-holes (BH), 
neutron stars (NS) and massive stars.

The wide range of characteristics prevents the systematic classification of GRBs.
Their energy spectrum is continuous and non-thermal,
and covers the range between $1$ keV and $1$ MeV;
their emission lasts from $10$ ms to $10^3$ s.
The Burst And Transient Source Experiment (BATSE) has detected GRBs
with an event rate of about one per day between 1991 and 2000.

One of the most important results of observations is that many GRBs are at cosmological distances\cite{meegan}.
This fact, recently confirmed also by the satellite BeppoSAX\cite{transX}, which has detected some optical
counterpart with redshifts close to $z=1$\cite{righ_gal1,righ_ass,righ_gal2,grb980703},
implies the emission of a
huge amount of electromagnetic radiation: in few seconds the GRBs release an energy of
$10^{51} - 10^{54}$ erg. On the other hand, this implies that the GRBs are rare
events in a galaxy: in fact, as the rate of GRB events is of the order $ 1/\text{day} $,
their rate amounts to $1$ event$/10^{6}$ years per galaxy\cite{piran}.

The theoretical framework which gives a coherent explanation of the experimental data set of GRBs is the
so-called ``\emph{fireball}'' model: an ``inner engine'' produces a flux of relativistic
energy of the order $10^{52}$ erg and it is extremely compact
and clearly not visible; this relativistic flux of particles is a kind of ``fireball'' able
to produce electromagnetic radiation in a optically  thin shell of matter. The inner engine
works for 1-3 days, producing the afterglow, but the bulk
of the explosion lasts typically $10$ s (with peaks around $0.5$ and $30$ s 
and a variation of six orders of magnitude, $10^{-3} \div 10^{3}$ s) when
the major part of the energy is emitted.


The inner engine progenitors are likely to be coalescing binary systems,
such as NS-BH or NS-NS systems,
or single stars that collapse into BH in supernova-like 
events (the so-called ``\emph{collapsar}'' models \cite{woosley_9912484}).
If this is the correct explanation of the GRBs, then we would have compact 
astrophysical systems for which gravity plays an important role,
and gravitational waves would be emitted\cite{grb_og,meszaros2,woosley_0007176}. 

This scenario has motivated us to investigate the behavior
of the gravitational wave detector AURIGA during time spans which include
the arrival time of a GRB burst.
The AURIGA detector \cite{auriga_amaldi} is an Al$5056$ resonant bar of about $2.3$ \emph{tons}
with a typical noise temperature of \hbox{$7$ mK}  
and a bandwidth of about  $1$ Hz. The detector is sensitive to gravitational waves (g.w.) signals
over 1 Hz bandwidth around each one of its two resonant frequencies i.e. 913 and 931 Hz.
The sensitivity of a g.w. detector can be conveniently expressed by the quantity
$h_{min}$, representing the minimum g.w. amplitude detectable at $ SNR
= 1$, which for the AURIGA detector is $h_{min}\sim 2\div 5 \times 10^{-19}$.
The AURIGA sensitivity
is enough to detect NS-NS, BH-NS and BH-BH mergers which take place within the Galaxy. 

Previous work 
on GRB-GWB association appeared in the literature,
specifically about a single GRB trigger \cite{bonifazi},
and about a set of GRB triggers
\cite{astone_99,modestino_vulcano98_finale,finn_pr};
here we present an experimental
upper limit on such an association obtained with the AURIGA-BATSE data.
 
The plan of the paper is as follows:
in {\rm Sec. II}  we discuss the two methods, coincidence and statistical searches, which we use for the
association of the GRBs in the BATSE catalog with GWBs. The results obtained with the two methods
are presented in {\rm Sec. III}. In  {\rm Sec. IV} we analize our results and discuss the possibilities
opened by our methods for future searches.

\section{Search methods}

The aim of our analysis is the search of an association between GRB and GWB arrival times
within a time window $W$.
The analysis has been carried out by means of two different procedures which have been already discussed
in the literature:
$i$) the correlation method\cite{igec} which is based on the coincidence between the arrival time of
candidate gravitational events selected over a given threshold and the GRB trigger
and $ii$) a statistical method \cite{finn_pr} which relies on a hypothesis test on a statistical
variable representing the mean
energy of the gravitational detector at the  GRB trigger.

The coincidence window plays an important role in both the analysis.
In fact the coincidence window should be
wide enough to hold the delay distribution between GRB and GWB. We notice that its
``optimal'' value depends on the astrophysical sources and on the g.w. detector properties.
If we assume that the GRBs are generated by internal shocks in the fireball, 
the delay between GRB and GWB is less than $1$ second
\cite{meszaros2} but there are still some uncertainties. Moreover, the delay is widen
by the cosmological redshift.  To be as much conservative as possible on the distribution
of the delays we choose $W=5$ s. This value turns out to be also consistent with the
requirements of the filtering procedures of the AURIGA detector.
In fact, the search for g.w. bursts requires a filtering of the data by
a Wiener-Kolmogorov filter matched to $\delta$-like signals \cite{analisi_ortolan_amaldi} 
and its characteristic time
is the inverse of the detector bandwidth  i.e. $\sim 1$ s. This time, as we shall see below,
establishes the timescale of the noise correlation of the filtered data. 

In what follows, ``$\delta$-like signal'' means any g.w. signal
which shows a nearly flat Fourier transform at the resonance frequencies of the
detector ($913$ and $931$ Hz) over a $\sim 1$ Hz bandwidth. Therefore the
metric perturbations $h(t)$ sensed by the AURIGA detector
can be a large class of short signals (of millisecond duration) including
the latest stable orbits of inspiralling NS-NS or NS-BH, the subsequent merging and the
final ringdown \cite{flan}
and the collapsar bursts \cite{woosley_0007176}
which could be expected signals associated with GRBs.

\subsection{Coincidence search}

The coincidence method has been successfully applied to the search of coincident
excitations of different g.w. detectors \cite{coinc-storico,prl}; much work has been devoted to exploit its
potentialities  and to develop robust estimates of the background of
accidental, even in the presence of non stationary event rates \cite{igec,prl}.
A g.w. candidate event is a local maximum of the filtered data corresponding to any
excitation of the detector.
From the AURIGA filtered output we extract a list of candidate events setting
an adaptive threshold in their signal-to-noise ratio ($SNR=5$).
The event lists used in this analysis belong to periods of satisfactory performance of AURIGA
as described in Ref. \cite{baggio_xx}. 
It is worth noticing that the AURIGA event search checks each event
against the expected signal template by means of a $\chi^2$ test \cite{chi2}.
The event lists contain the information needed
to describe a $\delta$-like signal namely, its time of arrival (in UTC units), the amplitude of the
Fourier transform, and the detector noise level at that time.
The BATSE data are taken from the $\mathrm{4B}$ catalog by Meegan et al.
available on Internet\cite{meegan_web}
and includes trigger time (in UTC units), right ascension and declination,
error box and other information about every GRB triggered.

We label with $t_A^{(i)}$ the estimate arrival time of the $i$-th
candidate g.w. event detected by AURIGA
and  with $t_\gamma^{(k)}$ the trigger time of the $k$-th
GRB detected by BATSE.
A coincidence between the $i$-th AURIGA event and the $k$-th GRB trigger time
is observed if $|t_A^{(i)}-t_\gamma^{(k)}|\leq W$, where $W$ is the coincidence window.

The coincidences found have to be compared with the accidental coincidence background
due to chance. Two standard methods to evaluate the probability of accidentals have been applied
to the pair BATSE-AURIGA: $i$) performing thousands time-shifts of the arrival time of one detector
with respect to the other and looking for accidental coincidences at each shift\cite{coinc-storico}; 
$ii$) assuming
independent Poisson distribution of event times and using the mean measured rates to estimate
the accidental rate $n_a$
\begin{equation}
n_a=N_A N_\gamma \frac{ \Delta T  }{ T } \ , \label{acc2}
\end{equation}
where $N_A$ is the number of AURIGA events in the period $T$,
$N_\gamma$ the number of GRB events in the same period
and $\Delta T= 2 W$ is the total amplitude of the coincidence window
(see Appendix \ref{app1} for a proof of this basic relation).
To avoid the problem of multiple coincidence within the same window
we prefer to estimate the background with $W=1$ s and to scale
our results with the help of Eq. (1) to $W=5$ s. The consistency
of the two estimates ensures that the arrival times of the AURIGA-BATSE pair are
distributed as Poisson random points even if the AURIGA event rate  has
been found to be not stationary\cite{igec,prl}.

\subsection{Statistical search}

The method of the statistical search has  been proposed for a pair of interferometric detectors
by L.~S. Finn and coworkers \cite{finn_pr};
here we slightly modify their approach to the case of a single resonant detector.
The data we use are the AURIGA filtered data obtained by means of the Wiener filter matched to
a $\delta$-like signal, without setting any threshold on their amplitudes.
If a signal enters the detector at time $t_0$, its output can be written as
\eq{y(t)=h f(t-t_0)+\eta(t) \ ,}
where $f(t-t_0)$ is the normalized signal template of the Wiener filter,
$h$ its amplitude (i.e. $f(0)=1$), $t_0$ its arrival time and
$\eta(t)$ is a stochastic process with zero mean and correlation
\eq{\mean{\eta(t),\eta(t')} = \sigma_h^2\, f(t-t') .}
Here $ \sigma_h^2$ is the variance of the filter output
in the absence of any signal and $f(t)$  is a superposition of two exponentially
damped oscillating functions \cite{timing} 
which can be approximately expressed as 
\eq{f(t) \approx  e^{-|t|/t_w} \cos(\omega_0 t) \cos(\omega_B t) \label{eq_ft} \ \ \ \ (\omega_0 t_w\gg 1)  \ , }
where
$t_w$ is the Wiener filter decaying time (i.e. the inverse of the detector bandwidth), $\omega_0$
a center carrier frequency and $ \omega_B$ is an amplitude
modulation frequency.
Typical values for the AURIGA detector are $t_w\simeq 1$ s, $\omega_0 \simeq 920$ Hz and $ \omega_B \simeq 20$ Hz.

Let us define now the random variable $X$, which represents an averaged measure of the energy
released by a g.w. signal impinging on the detector at time $t_0$
\eq{X(t_0)=\frac{1}{2 W} \int_{-W}^{W} dt\, |y(t-t_0)|^2 \ ,}
where $t_0$ is the center of the time window.
If an association between GRB and gravitational waves exists, 
the filtered output 
of the gravitational waves detector, in periods just prior
the GRB (``{\it \lon-source}'' population) will differ (statistically) 
from the output 
at other times (``{\it \loff-source}'' population).

A statistically significant difference between {\it \lon-} and  {\it \loff-source}
populations clearly supports a GWB-GRB association. 
For each  of the \non GRB trigger we compute $X(t_\gamma^{(k)})$, $k=1,\dots,N_{\lon}$
which forms the $\chi_{\lon}$ set of {\it \lon-source} events,
and construct a complementary set $\chi_{\loff}$ with \noff  {\it \loff-source} events
using windows before and after the trigger.
The sets $\chi_{\lon}$  and $\chi_{\loff}$ are samples drawn from the
populations whose distributions we denote  $p_{\lon}$  and $p_{\loff}$.
The {\it \loff-source} events are taken in periods not correlated with the trigger time,
at a distance in time greater than $10^3$ seconds from GRB the trigger, both before and after it; 
this should be sufficient to have a fair sample of the {\it \loff-source} events
as any GRB-GWB association is reasonably excluded.

For windows $W$ greater than the Wiener filter characteristic time,
the central limit theorem implies that  $p_{\loff}$ is a normal distribution.

Now suppose that the GWBs fall within the window $W$ opened around the GRB trigger time and
that the SNR
of the gravitational signal associated with the GRB 
(averaged over the source population) is smaller than one, then
$p_{\lon}$ is also a normal distribution with mean
\begin{eqnarray}
\mu_{\lon} &=& \mu_{\loff} + {\rm E}\left[ \frac{1}{2 W} \int_{-W}^{W} dt\, |h f(t-t_\gamma)|^2 \right] \nonumber \\
         &\simeq& \mu_{\loff} + \left(\frac{t_w}{8 W}\right) {\rm E}[h^2] \ \ \  (W\gg t_w) \ ,
\end{eqnarray}
where 
${\rm E}\left[\,\cdot\,\right]$ is the average over the astrophysical source population of GRBs. 

The basic idea behind the statistical approach is a hypothesis testing where
the null hypothesis ${\cal H}_0$ to test is the equivalence of the   {\it \loff-source} and  {\it \lon-source}
distributions:
\eq{ {\cal H}_0 : \hspace{0.8 cm} p_{\loff}(X) = p_{\lon}(X) .\label{test_1}}
The rejection of ${\cal H}_0$ clearly supports a GWB-GRB association.
Since  $p_{\lon}$  and $p_{\loff}$ are normal and could differ
only in their mean values, we can test  ${\cal H}_0$ by the Student's $t$-test\cite{student}.

The $t$ statistic is defined from $\chi_{\lon}$  and $\chi_{\loff}$ by
\begin{eqnarray}
t & = & \frac{\hat{\mu}_{\lon}-\hat{\mu}_{\loff}}{\Sigma}
\sqrt{ \frac{ N_{\lon} N_{\loff}}{ N_{\lon}+N_{\loff} } } \\
\Sigma^2 & = & \frac{(N_{\lon}-1)\hat{\sigma}_{\lon}^2 +
(N_{\loff}-1)\hat{\sigma}_{\loff}^2 }{N_{\lon}+N_{\loff}-2} \ ,
\end{eqnarray}
where $\hat{\mu}_{\lon}$ and $\hat{\mu}_{\loff}$ ($\hat{\sigma}_{\lon}^2$ and $\hat{\sigma}_{\loff}^2$)
are the sample means (variances) of $\chi_{\lon}$ and $\chi_{\loff}$, respectively.

The expected value of $t$ averaged on the source population and on the filtered output of
the detector is
\eq{\mu_t = {\rm E}[t] = \left(\frac{t_w}{8 W}\right) \frac{ {\rm E}[h^2] }{\sigma}
\sqrt{ \frac{ N_{\lon} N_{\loff}}{ N_{\lon}+N_{\loff} } } \ ,
 \label{def_mu_t} }
where $\sigma={\rm E}[\Sigma]$. 

Let us consider the upper limit on ${\rm E}[h^2]$
in the assumption  ${\cal H}_0$ is true: in this case the most probable value of ${\rm E}[h^2]$ is zero. 
From Eq. \ref{def_mu_t} we get
\begin{eqnarray} 
\left(\frac{t_w}{8 W}\right)  \frac{{\rm E}[h^2]}{\sigma} \leq \mu_{t,max} 
\sqrt{ \frac{ N_{\lon} + N_{\loff}}{ N_{\lon}N_{\loff} } } =  \nonumber \\
\null = \left\{
\begin{array}{ll}
\mu_{t,max} \sqrt{2/N_\gamma} & (N_{\lon}=N_{\loff}=N_\gamma) \\
\mu_{t,max} / \sqrt{N_{\lon}} & (N_{\loff}\gg N_{\lon})
\end{array}
\right.
\end{eqnarray}
where $\mu_{t,max}$ is the upper limit.
Assuming ${\cal H}_0$ true and $N_{\loff}\gg N_{\lon}$, we have
\eq{{\rm E}[h^2] \leq h^2_{max} = 
\left(\frac{8 W}{t_w}\right)
\frac{\mu_{t,max}}{\sqrt{N_{\lon}}}
\ \sigma\ ;}
the value of $\mu_{t,max}$ can be deduced by the selected confidence level. 
In this way we are able to set 
an upper limit on the average amplitude of gravitational signals associated with GRBs 
\begin{eqnarray}
h^2_{\mathrm{RMS}} &\leq&
\left[1.4\times 10^{-18}\right]^2 \:
\frac{W}{5\:s}
\left(\frac{t_w}{1\:s} \right)^{-1}
\frac{\mu_{t,max}}{1.96}
\times \nonumber \\
&\times& 
\left( \frac{N_{\lon}}{100}\right)^{-1/2}
\frac{\sigma}{[5\times10^{-19}]^2}.
\label{h_base}
\end{eqnarray}

\section{Results}

The AURIGA data used in the two analysis are relative
to the years 1997 and 1998, and the number of GRBs
which fall into the AURIGA data taking periods is 120.

\subsection{Coincidence search}

Within the window of 5 seconds we have found 2 events in coincidence. 
This experimental result has to be compared with 
the number of coincidences due to chance.

\begin{figure}[htb] 
\centering 
\leavevmode\epsfxsize=8.6cm
\epsfbox{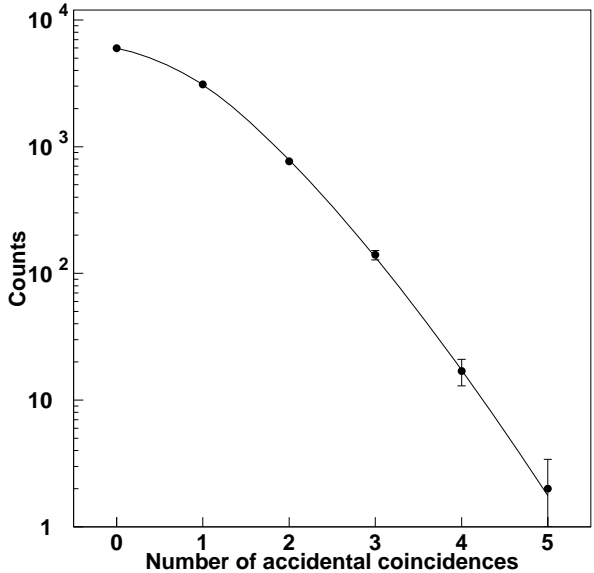}
\caption{
Poisson fit of the data obtained with the shifts method.
} \label{poiss}
\end{figure}

The shifts method consists of $10^4$ time shifts of 
the coincidence window; for each one we compute the number of coincidences;
if the candidate GWB arrival times and the GRB triggers
can both be modeled as Poisson random points\cite{papoulis},
the number of accidental coincidences is fitted to a Poisson curve;
from the fit we get the expected number of coincidences due to chance.
The results are summarized 
in Fig. \ref{poiss}.
From the fit we obtain a value of mean expected accidental coincidences of 
$n_{a}= 2.57 \pm 0.04 $.

Another approach to evaluate the number of accidental coincidences is given by Eq. \ref{acc2},
that holds in case of Poisson random points\cite{papoulis}.
The total number of AURIGA events and GRB triggers are respectively
$N_A=26816$, $N_\gamma=120$, assuming \hbox{$\Delta T = 2 W = 10$ s}
and \hbox{$T=1.32\times 10^7$ s} we get
$n_{a}=2.4\pm 0.2$.
The error on $n_a$ can be easily estimated assuming the Poisson statistics for
the fluctuations on $N_A$ and $N_\gamma$ i.e. $\sqrt{N_A}$ and $\sqrt{N_\gamma}$ respectively.
The two estimates of the accidental number of coincidences are
in good agreement and demonstrate that the two event rates are uncorrelated. 
We conclude that the 2 coincidence found are due to chance.

\subsection{Statistical search}

We have first tested the method using a Monte Carlo simulation by
adding $N_{on}$ signals at fixed SNR over a gaussian noise generated
by means of a noise model with the same parameters of the AURIGA detector.
The simulated detector output is   then fed to the same data filtering procedures
used for the AURIGA experimental data.
We have generated  120 $\delta$-like signals with SNR 3, 2 and 1  and 1000 with SNR 1
storing their true arrival times $t_0$.  The signals  have been superimposed to stationary
gaussian noise and we have then formed the $\lon$ and $\loff$
populations and calculated the $t$ value.   
The probability $P(t)$ that the $t$ value obtained is due to chance 
is reported in Table \ref{table_2}. 
For signal with $SNR=3$ and $SNR=2$ this probability is very small and the value 
of $t$ is statistically significant, showing a GRB-GWB association.
On the other hand, for signals with $SNR=1$, we have to increase the number of GRBs  to 1000
to get a statistical significance.

\begin{figure}[htb] 
\centering 
\leavevmode\epsfxsize=8.6cm
\epsfbox{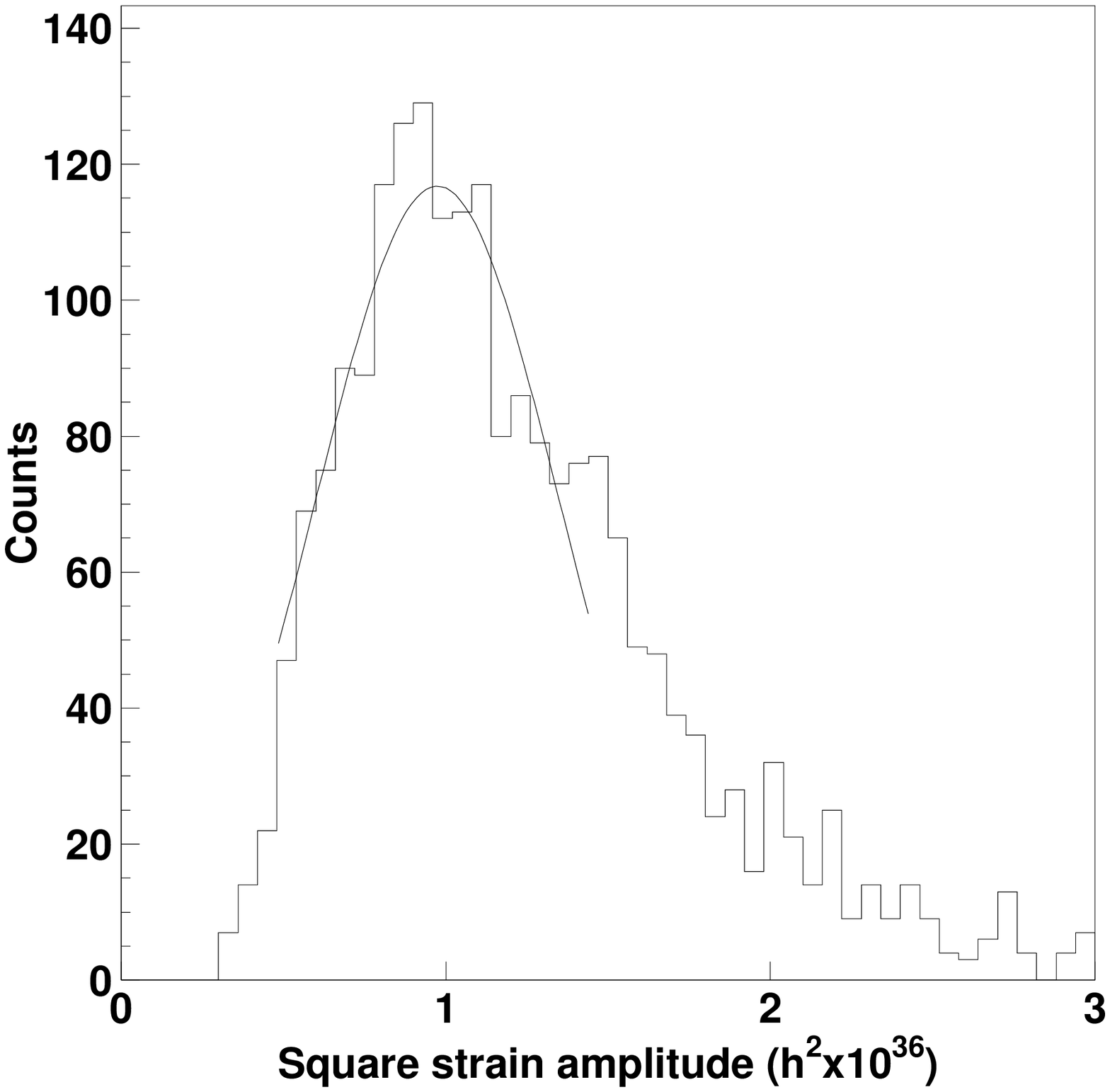}
\caption{
Gaussian fit of the \loff-source set of the statistical search.
} \label{off_data}
\end{figure}

\begin{figure}[htb] 
\centering 
\leavevmode\epsfxsize=8.6cm
\epsfbox{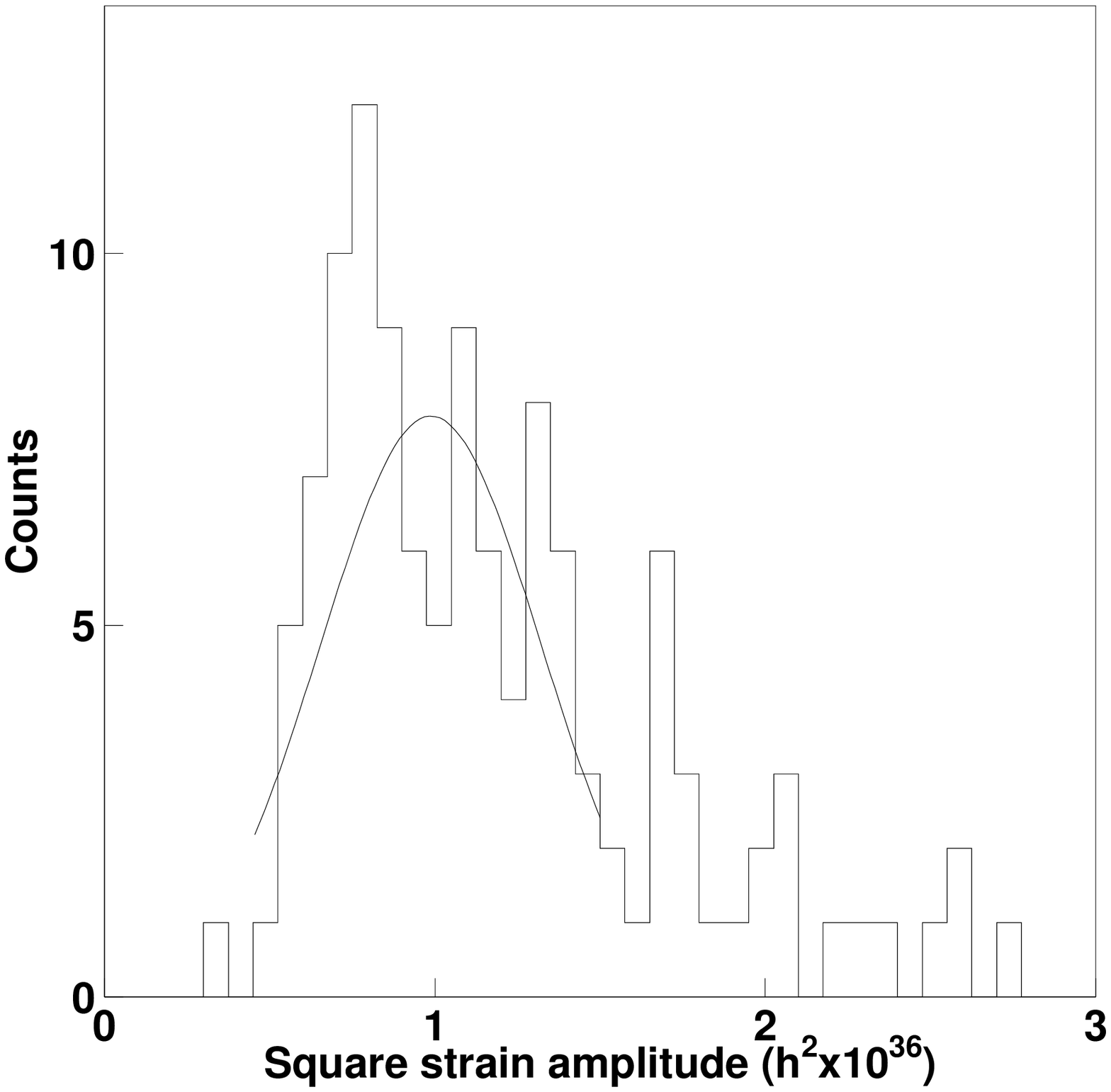}
\caption{
Gaussian fit of the \lon-source set of the statistical search.
} \label{on_data}
\end{figure}

The $\chi_{\loff}$ and $\chi_{\lon}$ sets
relative to the  AURIGA-BATSE data between 1997 and 1998
are given in the histograms of Fig. \ref{off_data} and Fig. \ref{on_data},
where the non gaussian tails are due to the non stationarity of the AURIGA noise.
The value of the Student's $t$-test obtained is $t=0.58$,
which corresponds to a probability of $0.28$ that it is due to chance.

As we conclude that there is no evidence of an association of GRB-GWB,
we can put a constraint on  gravitational radiation emitted from GRBs
averaged over the source population, $h_{\mathrm{RMS}}$.
The value of $\mu_{t,max}$ can be found setting a confidence level $\text{C.L.}$
and solving the equation:
\eq{\int_{-\infty}^{\mu_{t,max}} f_d(t)\,dt=\text{C.L.} \ , }
where $f_d(t)$ is the distribution function for the Student's $t$ with $d$ degrees of freedom.
Notice that for high $d$,  $f_d(t)$ tends to a normal curve with zero mean and unitary variance.
If we choose \hbox{C.L.=95\%} 
and set \hbox{$d=N_{\lon}+N_{\loff}-2$},
we get $\mu_{t,max}=1.65$.
Using Eq. \ref{h_base}, we set the following upper limit on the averaged g.w. amplitude
associated with GRBs:
\eq{h_{\mathrm{RMS}} \leq 1.5\times 10^{-18} .}

\section{Discussion}

The two search methods reported in this paper show
no evidence of a correlation between $\gamma$-ray bursts and gravitational waves:
First, the two coincidences found with the coincidence search have no statistical
significance and can be reconduced to chance.
Next, the statistical method doesn't lead to a statistically significative value of the Student's $t$.
However we were able to put an upper limit on gravitational signals associated to the GRBs
averaged over the source population, $h_{\text{RMS}} \leq 1.5 \times 10^{-18}$.  

The existence of burst-like excitations (usually referred as non-gaussian noise) in the AURIGA data is well known,
and here we deal with this noise selecting the periods of time when the detector was operating in a satisfactory way.
The vetoing procedure of data dominated by non gaussian noise
has been set by the AURIGA data analysis \cite{baggio_gwdaw} and the resulting duty cycle is
about \hbox{40 \%} 
of the total operating time.
Care was also taken to cope with non stationarities of the AURIGA noise
that give rise to the non gaussian tails in Fig. \ref{off_data} and Fig. \ref{on_data} excluding such tails
from the fits we used to estimate the statistical variable $t$.
However, it is important to notice that even with the above limitations the analysis has
reached a level of sensitivity which is astrophysically of some interest. 
In order to get better 
upper limits we have also explored 
the possibility of using in our analysis the incoming direction of GRBs but   
we have found no significant improvements (see Appendix \ref{app_B}).

To increase the confidence in our results we have applied 
the non-parametric Mann-Whitney $u$-test \cite{u-test} 
to the sample sets of the statistical search, obtaining a second confirmation of the null hypothesis.
Ranking the elements of the union set \hbox{$\chi_{\loff} \oplus \chi_{\lon}$} in increasing order,
we get a statistical parameter $z=1.59$, smaller than the critic value $z=1.95$
imposed by the fixed C.L. of 95 $\%$.

The upper limit we have set can be improved in the future simply by
the increasing of common data taking periods of AURIGA ($N_{on}$ increases) and the new experiment  
HETE-II (High Energy Transient Explorer \cite{hete2}) which is going to substitute by now the wasted BATSE satellite
in the GRB search. 
Another possibility to reach more astrophysically interesting sensitivities 
is the upgrade of the AURIGA detector which is now in progress\cite{zendri_sigrav}. The 
predicted sensitivity and bandwidth of the 
AURIGA detector equipped with the new read out system would be respectively  
$h_{min}\approx 8\times 10^{-20}$ and $t_w^{-1}\approx 10$ Hz. 
This sensitivity, together with an enhancement of noise stationarity 
and duty cycle of the detector, 
would correspond to the lowering of the upper limit 
$h_{\mathrm{RMS}}$ of about 2 order of magnitude in one year  
of correlation analysis.

\appendix 
\section{}
\label{app1}

To derive Eq. \ref{acc2} let us consider  $n$ points random distributed in a time interval $[0,T]$.
The probability $P\left(\{k_a,t_a\}\right)$ 
that $k_a$ points lies in the time window $t_a=t_2-t_1$ is given
by the binomial distribution \cite{papoulis}
with probability  $p=t_a/T$ that a single point lie in $t_a$.
If $n\gg 1$ and $t_a \ll T$, using the Poisson theorem 
we get
\eq{P\left(\{k_a,t_a\}\right) = e^{- n t_a / T } \frac{(n t_a / T )^{k_a}}{k_a!}.}
For $m$ not overlapping windows, it can be demonstrated that
in the limit of $n\rightarrow\infty$, $T\rightarrow\infty$ and $n/T$ constant,
the probability of \{{\it $k_1$ points in $t_1$}\}, \ldots, \{{\it$k_m$ points in $t_m$}\} 
is \cite{papoulis}
\begin{eqnarray}
P\left(\{k_1,t_1\},\dots,\{k_m,t_m\}\right) &=& \prod_{i=1}^{m} e^{- n t_i / T } 
\frac{( n t_i / T )^{k_i}}{k_i!} \nonumber \\
 &=&  \prod_{i=1}^{m}P\left(\{k_i,t_i\}\right) \label{P_multi}
\end{eqnarray}
showing that the events \{{\it$k_i$ points in $t_i$}\} and \{{\it$k_j$ points in $t_j$}\}
are independent for every $i$ and $j$.
Substituting $m=N_\gamma$, $t_i=2W=\Delta T$ and $n =  N_A$ we get
\begin{eqnarray}
&& P\left(\{k_1,\Delta T\},\dots,\{k_{N_\gamma},\Delta T\}\right) = \nonumber \\ 
&=& e^{- N_\gamma N_A \Delta T / T } \left( \frac{ N_A \Delta T } {T} \right)^k
\prod_{i=1}^{N_\gamma} \frac{1}{k_i!}.  
\end{eqnarray}

We can now evaluate the probability $P(k)$ to have $k$ coincidences,
that can be obtained  taking into account, at fixed $k$, all the possible sets 
$[k_i]=\{k_1,\cdots,k_{N_\gamma}\}$ 
with the constrain \hbox{$\sum_{i=1}^{N_\gamma} k_i \label{summ_k} = k $}; we get
%
\begin{eqnarray}
P(k) &=& 
\sum_{[k_i]} P\left(\{k_1,\Delta T\},\dots,\{k_{N_\gamma},\Delta T\}\right)  \nonumber \\
\null &=&  e^{- N_\gamma N_A \Delta T / T }  \left( \frac{ N_A \Delta T } {T} \right)^k 
\sum_{[k_i]}
\prod_{i=1}^{N_\gamma} \frac{1}{k_i!} \nonumber \\
\null &=&  e^{- N_\gamma N_A \Delta T / T }  \left( \frac{ N_A \Delta T } {T} \right)^k 
\frac{N_\gamma^k}{k!} \label{eq__a5} \ .
\end{eqnarray}
Rearranging the factors in Eq. \ref{eq__a5} 
we get a Poisson distribution
as in Eq. \ref{acc2}.
%
The last equality can be easily demonstrated using the multinomial 
expansion relation \cite{multinomial}
\eq{ \left(\sum_{i=1}^N x_i\right)^k  =  k!  \sum_{[k_i]} 
\prod_{i=1}^{N} \frac{x_i^{k_i}}{k_i!} \label{es_mult}}
and substituting $x_i=1$, $i=1\cdots N$.

\section{}
\label{app_B}

As the incoming direction of GRBs is known, one may wonder if a selection of the GRBs
based on a cutoff 
on the AURIGA antenna pattern, averaged on the unknown polarizations of the GWB, 
could increase the sensitivity of our analysis.

The antenna pattern of a resonant bar detector is given by
\eq{F(\theta) = 1- \left( {\bf \hat k}\cdot {\bf \hat z} \right)^2 = \sin^2(\theta)\ ,}
where ${\bf \hat k}$ and ${\bf \hat z}$ are unitary vectors parallel to the GRB direction and
the antenna bar axis and $\theta$ is the angle between ${\bf \hat k}$ and ${\bf \hat z}$.
Sources that fall outside the two cones, 
which are defined by the equation ${\bf \hat k}\cdot {\bf \hat z} \geq \cos(\xi)$, 
have a figure pattern $F(\theta)\geq\sin^2(\xi)$ and therefore the
average energy associated with these g.w. sources is
\begin{eqnarray} 
{\rm E^\xi}[h^2]
&\propto& \frac{15}{16} \int_{\Omega_\xi}
F^2(\theta) \ \frac{d\Omega}{4\pi} \nonumber \\
&=&\frac{75}{64}\left[ \sin(\xi) + \frac{1}{6} \sin(3\xi) + 
\frac{1}{50}\sin(5\xi)\right]\label{eqb1} \ ,
\end{eqnarray}
where the solid angle $\Omega_\xi$ is defined by $ F(\theta)\geq F(\xi)$.
Moreover, as GRBs are isotropically distributed over the sky, the cutoff on the figure pattern
decreases the number of available GRB: $N^\xi_{on} = N_{on} \sin(\xi)$.
Therefore the net effect of this cutoff on the expected value of $t$ in Eq. \ref{def_mu_t} is
\begin{eqnarray}
\mu_t^\xi
&=& 
\mu_t\frac{75}{64} \left[ \sin(\xi) + \frac{1}{6} \sin(3\xi) + 
\frac{1}{50}\sin(5\xi)\right] \times \nonumber \\
&\times& \sin(\xi)^{1/2}\label{eq_app2} \ .
\end{eqnarray}
The function $\mu_t^\xi/\mu_t$  
is a continuously increasing function in the range $\xi \in [0,\pi/2]$,
and  $\mu_t^\xi/\mu_t = 1$ at $\xi = \pi/2$ (i.e. the whole solid angle).
We must conclude that a cutoff on the GRB direction does not enhance 
the sensitivity  
of the statistical search.

\nocite{weber} 



\begin{table}
\caption{ 
Results of a Monte Carlo simulation of the statistical search.
The column $P(t)$ is the probability that the estimated $t$  is due to chance.
} \label{table_2}
\begin{tabular}{crcr}
SNR of the generated signals & \non & t & $P(t)$ \\
\tableline 
\\
3 & 120  & 8.1  & $ < 10^{-9}$ \\ 
2 & 120  & 3.4  & $4\times 10^{-4}$   \\
1 & 120  & 0.4  & $3\times 10^{-1}$   \\
1 & 1000 & 3.6  & $10^{-4}$         \\
\end{tabular}
\end{table}

\begin{table}
\caption{
Results of the fits on the $\chi_{\loff}$  and $\chi_{\lon}$ sets with a 
gaussian distribution. 
}\label{table_3}
\begin{tabular}{rccc}
& $N$ & $\mu$ & $\sigma$ \\
\hline 
\\
\loff & $2206$ & $(0.97\pm 0.02)\times 10^{-36}$ & $(0.37\pm 0.02)\times 10^{-36}$ \\
\lon  & $120$  & $(0.99\pm 0.06)\times 10^{-36}$ & $(0.33\pm 0.06)\times 10^{-36}$ \\
\end{tabular}
\end{table}



\end{document}